\documentclass[11pt]{article}
\pdfoutput=1
\usepackage[export]{adjustbox}
\usepackage{jheppub}
\usepackage{bbm}
\usepackage{tikz}
\usetikzlibrary{calc}
\usepackage{graphicx,subcaption}
\usepackage{mathtools}
\usepackage{bm}
\usepackage{enumitem}
\usepackage{verbatim}
\usepackage{amssymb}
\usepackage{slashed}
\usepackage{lscape}
\usepackage{cases}

\newcommand{\blank}[1]{}

\newcommand\scalemath[2]{\scalebox{#1}{\mbox{\ensuremath{\displaystyle #2}}}}

\newcommand{\cN}{{\cal N}}

\newcommand{\bfmtb}{\bfmtb}
\renewcommand{\bfmtb}{\lambda}

\DeclareMathAlphabet{\mathbfsf}{OT1}{cmss}{bx}{n}

\renewcommand{\=}{\;= \;}

\newcommand\be{\begin{equation}}
\newcommand\ee{\end{equation}}
\newcommand\beq{\begin{equation}}
\newcommand\eeq{\end{equation}}
\newcommand\bea{\begin{eqnarray}}
\newcommand\eea{\end{eqnarray}}

\newcommand{\ket}[1]{{|\,#1\,\rangle}}
\renewcommand{\tilde}{\widetilde}
\renewcommand{\hat}{\widehat}
\makeatletter
\newcommand{\hathat}[1]{% 
\begingroup%
  \let\macc@kerna\z@%
  \let\macc@kernb\z@%
  \let\macc@nucleus\@empty%
  \hat{\mathchoice%
    {\raisebox{.3ex}{\vphantom{\ensuremath{\displaystyle #1}}}}%
    {\raisebox{.3ex}{\vphantom{\ensuremath{\textstyle #1}}}}%
    {\raisebox{.3ex}{\vphantom{\ensuremath{\scriptstyle #1}}}}%
    {\raisebox{.2ex}{\vphantom{\ensuremath{\scriptscriptstyle #1}}}}%
    \smash{\hat{#1}}}%
\endgroup%
}

\newcommand{\tildetilde}[1]{% 
\begingroup%
  \let\macc@kerna\z@%
  \let\macc@kernb\z@%
  \let\macc@nucleus\@empty%
  \tilde{\mathchoice%
    {\raisebox{.3ex}{\vphantom{\ensuremath{\displaystyle #1}}}}%
    {\raisebox{.3ex}{\vphantom{\ensuremath{\textstyle #1}}}}%
    {\raisebox{.3ex}{\vphantom{\ensuremath{\scriptstyle #1}}}}%
    {\raisebox{.2ex}{\vphantom{\ensuremath{\scriptscriptstyle #1}}}}%
    \smash{\tilde{#1}}}%
\endgroup%
}

\makeatother

\newcommand{\hatX}{{X}}

\begin{document}

\title{Explicit expressions for Virasoro singular vectors}

\author{G\'erard M.~T.~Watts}
\affiliation{Department of Mathematics,\\King's College London,\\
Strand, London WC2R 2LS, United Kingdom}
 \emailAdd{gerard.watts@kcl.ac.uk}
 
\abstract{
We present two explicit expressions for generic singular vectors of type $(r,s)$ of the Virasoro algebra. These results follow from the paper of Bauer et al which presented recursive methods to construct the vectors. The expressions presented here generalise the results of Benoit-Saint Aubin for the type $(1,s)$ singular vectors in two different ways: the first simply solves the recursion through the use of partitions; the second gives explicit formulae for the coefficients in a particular expansion.
A Mathematica notebook is available which implements the formulae.}

\maketitle

\section{Introduction}

The presence of singular vectors in representations of the Virasoro algebra is, of course, very well known and can be considered key to many of the properties of Virasoro minimal models in particular, leading to constraints on the fusion algebra \cite{FEIGIN1988209}, differential equations for correlation functions \cite{BELAVIN1984333}, modular differential equations for characters \cite{MATHUR1988303} and most recently appearing in decompositions of the identity \cite{fortin2024virasorocompletenessrelationinverse}. 

Very briefly, in each Verma module of type $M_{(r,s)}$ there is a singular vector at level $rs$ for $r,s$ non-negative integers. 
The precise form of the vectors is often not necessary, but despite this, there has been a lot of interest in their calculation.
The first results concerned the $(1,s)$ singular vectors, where Benoit and Saint Aubin \cite{BENOIT1988517} (BSA) provided explicit expressions of the singular vectors involving a sum over partitions.  
This formula was ``explained'' by Bauer et al \cite{Bauer:1991qm,BAUER1991515} as the solution of a recursion relation coming from fusion of specific primary fields and a recursive method to find the generic $(r,s)$ singular vectors was outlined in proposition 5.2 in \cite{BAUER1991515}.

Various other methods to find expressions of the $(r,s)$ singular vectors are also known - an analytic continuation method by Kent \cite{Kent_1991}, through quantum Drinfeld-Sokolov reduction by Ganchev and Petkova \cite{Ganchev_1992,Ganchev_1993}, and through the field realisations \cite{Ridout:2014zfa,Mimachi,Sakamoto_2005}, each of which can be helpful in particular circumstances. 
The method of Bauer et al was also extended to singular vectors of type $(2,s)$ by Millionshchikov \cite{Millionshchikov}.

It has often been stated that there is no explicit formula for the generic $(r,s)$ singular vectors, but it is explained here such a formula can be found simply by ``solving'' the recursion of \cite{BAUER1991515} in exactly the same way the the BSA expressions can be found from a simpler recursion by a sum over partitions, to give our first explicit expression.
Having found such an expression, it is only a little more work to extract the coefficients of the monomials $L_{-m_1}\ldots L_{-m_n}$ to give a second formulation for the singular vectors of type $(r,s)$. These two expressions are summarised in sections \ref{sec:f1} and \ref{sec:f2}. 

%This note explains how the recursion of Bauer et al (Proposition 5.2 in \cite{BAUER1991515}) can be solved in general to find explicit formulae as repeated sums over partitions. The first formulation \eqref{recur4} is little more then rewriting the recursion relation; the second uses this to explicitly find the coefficients in a particular form \eqref{eq:zeta1}, the same form as the BSA vectors.

The paper is structure as follows.
We first review the notation in section \ref{sec:2} before summarising the main results in sections \ref{sec:f1} and \ref{sec:f2}. We then
explain the recursion and its solution in \ref{sec:3}, and explain how to extract the required coefficients in \ref{sec:5}.
Finally we make some comments in section \ref{sec:conc}

\section{Notation and summary of results}
\label{sec:2}

The Virasoro algebra has generators $L_m$, central element $c$ and commutation relations 
\be
{}[L_m,L_n]
 = \frac{c}{12} m(m^2-1)\delta_{m+n,0} + (m-n)\,L_{m+n}
\ee
A highest weight state of the Virasoro algebra of weight $h$, $\ket h$, is a state that satisfies
\be
L_m \,\ket h = 0 \;,\; m>0
\;,\;\;\;\;
L_0 \,\ket h = h \,\ket h
\;.
\ee
The Verma module with highest weight $h$, $V_h$ is the module spanned by the states
\be
L_{-m_1}L_{-m_2}\cdots L_{-m_n}\ket h
\;,\;\;
m_1 \geq m_2 \geq \ldots \geq m_n>0
\;,
\label{eq:Vermabasis}
\ee
and the subspace of states at level $n$ is spanned by the states \eqref{eq:Vermabasis} with $\sum m_i = n$. The module
$V_h$ 
%This Verma module $V_h$ with highest weight state $\ket h$ 
contains a highest weight state (singular vector) $\ket{\psi_{r,s}}$ at level $rs$ whenever
$c=c(t), h=h_{r,s}(t)$, where $r,s$ are non-negative integers and 
\be
c(t) = 13 - 6 t - \frac 6 t
\;,\;\;
h_{r,s}(t) = 
(r^2-1)\frac t4 + (s^2-1)\frac 1{4t} - \frac{rs-1}2\;.
\ee
It is known \cite{Fuchs} that the singular vector $\ket{\psi_{r,s}}$ can be normalised as %and put 
%in the form
\be
\ket{\psi_{r,s}} = \cN_{r,s}\,\ket{h_{r,s}}
\;,\;\;
{\cN}_{r,s}
%\,\ket{h_{r,s}}
%\ket\psi
= %\left( 
(L_{-1})^{rs} + \sum_{m_1\geq..\geq m_r}c(m_1,\ldots,m_r)\,L_{-m_1}..L_{-m_r}
%\right)
%\ket{h_{r,s}}
\;.
\label{eq:psinorm}
\ee
%We denote this state at level $rs$ by ${\cN}_{r,s}\,\ket{h_{r,s}} $ where  ${\cN}_{r,s}$ is a polynomial in the ``lowering mode'' $L_m, m<0$. 
The purpose of this paper is to give an explicit expressions for ${\cal N}_{r,s}$ as a function of $t$.

We will find it useful to have some notation for sequences of integers. We will denote sequences by bold face letters (eg $\bm m$) and the length of a sequence by $|\cdot|$, for example
\be
\bm m = (m_1,m_2,\ldots ,m_l)
\;,\;\;
 l= |\bm m|
 \;.
\ee
It will be helpful to introduce notation for various partial sums of the terms
which we denote $\tilde m_i$, $\hat m_i$ and $\hathat m_i$,
\be
\tilde m_i = \sum_{j=i+1}^{|\bm m|} m_j
\;,\;\;
\hat m_i = \sum_{j=1}^{i} m_j
\;,\;\;
\hathat m_i = \sum_{j=1}^{i-1} m_j = \hat m_i - m_i
\;.
\label{eq:sumsdef}
\ee
as in this example,
\begin{align}
\bm m &= (m_1,m_2,m_3)
\;,\;\;
\\
\tilde{\bm m} &=  
(\tilde m_1,\tilde m_2,\tilde m_3)
= (m_2 + m_3, m_3, 0)
\;,\;\;
\\
\hat{\bm m} &=  
(\hat m_1,\hat m_2,\hat m_3)
= (m_1,m_1+m_2,m_1+m_2 + m_3)
\;,\;\;
\\
\hathat{\bm m} &= 
  (\hathat m_1,\hathat m_2,\hathat m_3)
= (0,m_1,m_1+m_2)
\;.
\label{eq:setdefs}
\end{align}

Particular sequences we need are the unordered partitions of the integer $n$ into positive integers, denoted by $\Pi_n$, and the unordered partitions of the integer $n$ into $r$ non-negative integers, denoted by $\Pi_{n,r}$. By ``unordered partition'', we mean a partition in which the elements are not ordered, so that the order is important, so that, for example,
\be 
\Pi_2 = \{ (2),(1,1)\}
\;,\;\;
\Pi_{2,3} = 
\{ (2,0,0), (0,2,0), (0,0,2), (1,1,0), (1,0,1), (0,1,1)\}
\;.
\ee
The partitions $\bm m$ defined above are also equivalent to ordered sets of integers by $i_j = \hat m_j$, so that 
$1 \leq i_1 < i_2 \ldots < i_r = n$ for $\bm m\in \Pi_n$ (with $r\geq 1$) and 
$0 \leq i_1 \leq i_2 \ldots \leq i_r = n$ for $\bm m\in \Pi_{n,r}$, and so all sums over $\Pi_n$ and $\Pi_{n,r}$ can be written in this form, which may be more familiar, especially for $\Pi_{n,r}$

We will also need the coefficients $\beta_s(\bm m)$ that appear in the BSA expression \eqref{eq:BSA},
\be
\beta_s(\bm m) = \prod_{j=1}^{|\bm m|-1}
\frac{t}{\tilde m_j(\tilde m_j-s)}
\;,
\label{eq:betadef}
\ee
and the factors $\gamma_r(p,s)$ which were found in \cite{BAUER1991515},
\be
\gamma_r(p,s) = \frac{r(p(s-1)-r)}{(s-1)!^2} 
\,\prod_{k=1}^{s-2}(kpt - rt - k(s-k-1))
\label{eq:gammadef}
\ee

Finally, for any sequence $\bm m$, of length $l$, we define the sequence $-\bm m$ as
\be
-\bm m = (-m_1,\ldots,-m_l)
\;,
\ee
and the corresponding monomials $L_{\bm m}$ and $L_{-\bm m}$ by
\be
L_{\bm m} = L_{m_1}L_{m_2}\cdots L_{m_l}
\;,\;\;
L_{-\bm m} = L_{-m_1}L_{-m_2}\cdots L_{-m_l}
\;.
\label{eq:mono}
\ee

We are now in a position to state the two results of the paper.

\subsection{First formulation of $\cN_{p,s-1}$}
\label{sec:f1}

Owing to the details of the construction, it is simpler to give the expressions for $\cN_{p,s-1}$ (as opposed to $\cN_{p,s}$),
\begin{align}
\cN_{p,s-1}
=&
\frac{1}{\zeta_{p,s}}
\sum_{\bm\pi\in \Pi_{p(s-1)}}
%(-1)^{|\bm\pi| -1}
\left(
\prod_{i=1}^{|\bm\pi|-1}
\frac{-1}{\gamma_{\tilde \pi_i}(p,s)}\right)
\cdot 
\left(
\prod_{i=1}^{|\bm\pi|}
\hatX_{\pi_i,\tilde \pi_i}
\right)
\;.
\label{eq:f1}
\end{align}
The factors $\gamma_r(p,s)$ are given in \eqref{eq:gammadef}, and the operators $X_{\pi,\tilde\pi}$ by
\begin{align}
\hatX_{\pi,\tilde\pi}
 = \sum_{\bm m \in \Pi_s}
 \beta_s(\bm m)
 \sum_{\bm k \in \Pi_{\pi,|\bm m|}}
 \prod_{i=1}^{|\bm m|}
 S^{\tilde \pi,\tilde m_i,\tilde k_i}_{m_i,k_i}
 \;,
 \label{xdef}
\end{align}
where $\beta_s(\bm m)$ is defined in \eqref{eq:betadef} and
%where we have no longer suppressed the dependence of the terms $S_{m,k}$ on $\tilde\pi,\tilde m_i$ and $\tilde k_i$
\begin{numcases}{S^{\tilde\pi,\tilde m,\tilde k}_{m,k} = }
(-1)^m(mh_{p,0} -h_{p,s-1} - h_{1,s} + \tilde m - \tilde\pi - \tilde k)
 \;,\;\;
 & 
 $k=0\;,$ 
\label{eq:Smo}
\\[2mm]
 0 \;,\;\; & 
 $k>0,m=1 \;,$
 \\[2mm]
 \binom{k-2}{m-2}
 L_{-k}\;, 
 & $k>0,m>1\;.$
\label{eq:Smk}
\end{numcases} 
%\end{align}
  %
\blank{
\begin{align}
&S^{\tilde\pi,\tilde m,\tilde k}_{m,0} = (-1)^m(mh_{p,0} -h_{p,s-1} - h_{1,s} + \tilde m - \tilde\pi - \tilde k)
 \;,\;\;
 \label{eq:Smo2}
 \\
&S^{\tilde\pi,\tilde m,\tilde k}_{m,k} = \begin{cases}
    \left({\small\begin{array}{c} k{-}2 \\ m{-}2 \end{array}}\right)  L_{-k} & m>1,k\geq 1 \;,\\
    0 & m=1,k\geq 1 \;.
\end{cases} 
\label{eq:Smk2}
\end{align}
}
Finally, the normalisation factor $\zeta_{p,s}$ is given by 
\be
\zeta_{p,s} = \zeta( (1,1,\ldots,1))_{p,s}
\;,
\ee
which is defined in equation \eqref{eq:zeta2}.

\subsection{Second formulation of $\cN_{p,s-1}$}
\label{sec:f2}

If we expand the solution \eqref{eq:f1} over the monomials $L_{-\bm
  m}$ (defined in \eqref{eq:mono}), we find the following expression for
the operator $\cN_{p,s-1}$ and 
the coefficients $\zeta(\bm m)_{p,s}$,
\begin{align}
&\cN_{p,s-1} = \frac{1}{\zeta_{p,s}}
\sum_{\bm m\in\Pi_{p(s-1)}}{\zeta{(\bm m)_{p,s}}} \;
L_{-\bm m}\;,
\label{eq:f2}\\
 \blank{   & \zeta(\bm m)_{p,s}   =
    \sum_{{\bf k}\in \Pi_{|{\bm m|}}}
    \Biggl( \prod_{j=2}^{|\bm k|} 
    \frac{-1}{\gamma_{{\tilde m}_{{\hathat k}_j}}\!(p,s)}
    \Biggr)
    \Bigg( \prod_{j=1}^{|\bm k|}
    \Bigg(\sum_{\bm q \in \Pi_s} \beta_s(\bm q)
    \nonumber\\ &
    \sum_{\bm v \in \Pi_{|\bm q|-k_j,k_j+1}}
    \Biggl(
    \prod_{w=1}^{k_j}
    S_0(q_{\hat v_w + w},m_{{\hathat k}_j+w})
% CHECK
    \Biggr)
    \Biggl(
    \prod_{w=1}^{k_j+1}
    \prod_{x=1}^{v_w}
    S_1(q_{\hathat v_w + w + x - 1},\tilde q_{\hathat v_w + w + x - 1},
    \tilde m_{\hathat k _j + w - 1})
    \Biggr)
    \Biggr)\Biggr) 
}
& \zeta(\bm m)_{p,s}   =
    \sum_{{\bf k}\in \Pi_{|{\bm m|}}}
    \Biggl( \prod_{j=2}^{|\bm k|} 
    \frac{-1}{\gamma_{{\tilde m}_{{\hathat k}_j}}\!(p,s)}
    \Biggr)
    \Biggl( \prod_{j=1}^{|\bm k|}
    Y(\bm m^j,\tilde m_{\hat k_j})
    \Biggr) 
    \label{eq:zeta2}
\end{align}    
where
\begin{align}
&Y(\bm m^j,\tilde m_{\hat k_j})
= 
\sum_{\bm q \in \Pi_s}
\beta_s(\bm q)
\Biggl[
\nonumber\\
&
\sum_{\bm v \in \Pi_{|\bm q| - k_j,k_j}}
\left(\prod_{w=1}^{k_j}
S_0(q_{\hat v_w + w},m^j_w))\right)
\left(
\prod_{w=1}^{k_j+1}
\prod_{x=1}^{v_w}
S_1( q_{\hathat v_w +w + x -1} ,
{\tilde q}_{\hathat v_w + w _ x -1},
{\tilde m}^j_{w-1} + {\tilde m}_{\hat k_j})
\right)\Biggr]
\;,
\label{eq:Ydef1}
\end{align}
\begin{align}
&\bm m^j = (m_{\hathat k_j+1},\ldots,m_{\hat k_j})
\;,\;\;
k_j = |\bm m^j|
\;,\;\;
%\\
%&
\zeta_{p,s}  = 
\zeta((1,1,\ldots,1))_{p,s}
\;,
\label{eq:zetaps}
\end{align}
$\beta_s(\bm q)$ is defined in \eqref{eq:betadef} and
%
%\begin{align}
%S(r,k,m) &= (-1)^r( k + (r-1) h_{p,0} - y - m)
%\;,\\
%S(0,k,m) &= \begin{cases} 0 & r=1 \\
% \left({\small\begin{array}{c} k{-}2 \\ r{-}2 \end{array}}\right) & r \geq 2 \end{cases}
%\end{align}
\begin{numcases}{S_0(m,k)= }
 0 \;,\;\; & 
 $m=1 \;,$
 \\[2mm]
 \binom{k-2}{m-2}
 \;, 
 & $m>1\;.$
\label{eq:S2b}
\end{numcases} 
\begin{align}{S_1(r,k,m)= }
(-1)^r(k+rh_{p,0}+h_{1,s}-h_{p,s-1} - m)
 \;.\;\;
 & 
 \label{eq:S2a}
\end{align} 
Note that this is a sum over all monomials $L_{-\bm m}$ with $m_i>0$, and, in exactly the same manner as for the BSA expressions, this is an over-complete set of monomials for all cases except the highest weight operators $\cN_{1,1}$, $\cN_{1,2}$ and $\cN_{2,1}$ and so the expressions for $\zeta({\bm m})_{p,s}$ given here are just one of an infinite set of possible expressions for the coefficients in \eqref{eq:zeta2}.

\section{The general form of the recursion relation for $\cN_{p,s-1}$}
\label{sec:3}

The idea behind the recursive definition of Bauer et al
\cite{Bauer:1991qm,BAUER1991515} is to use the following fusion of
Virasoro primary fields: 
\be
 [1,s] \times [p,0] \to [p,s{-}1]
 \;.
 \ee
 In this way, one can construct a sequence of states $\ket{f_n}$ in the Verma module $M_{h_{p,s-1}}$ as the action of the chiral field $\phi_{1,s}(z)$ on the highest weight state $\ket{h_{p,0}}$:
 \be
 \phi_{1,s}(z)\,\ket{h_{p,0}}
 = \sum_{m=0}^\infty \,z^{y+m}\, \ket{f_m}
 \;,
 \ee
 where 
 \be
 \ket{f_0} = \ket{h_{p,s-1}}
 \;,\;\;
 y=h_{p,s-1} - h_{p,0} - h_{1,s}
 = \frac{1}{4t} -\frac12(ps - p - s + 2)
 \;.
 \ee
 One can also consider the field 
 ${\hat\cN}_{1,s}\phi_{1,s}(z)$ corresponding to the state
 $\cN_{1,s}\ket{h_{1,s}}$ and a second sequence of states $\ket{\tilde
   f_i}$ in $M_{h_{p,s-1}}$, 
 \be
 {\hat N}_{1,s}\phi_{1,s}(z)\,\ket{h_{p,0}}
 = \sum_{i=0}^\infty z^{y+s+i}\,\ket{\tilde f_i}
 \;.
 \ee
 By using the standard relations for the action of a mode $\hat L_m$ on a field, \eqref{eq:Lmode1} and \eqref{eq:Lmode2}, this can also be expressed as
 \be
 {\hat N}_{1,s}\phi_{1,s}(z)\,\ket{h_{p,0}}
 = \sum_{i=0}^\infty z^{y+s+i}\,\sum_{r=0}^i\,
 \hatX_{r,i-r}^{p,s}\ket{f_{i-r}}
 \;,
 \ee
 where $X_{m,n}^{p,s}$ are combinations of lowering modes of the Virasoro algebra which can be derived from the expression for $\cN_{1,s}\ket{h_{1,s}}$.
 As a consequence, one has the relations
 \be
 \ket{\tilde f_i}
 = \sum_{r=0}^i \hatX_{r,i-r}^{p,s}\,\ket{f_{i-r}}
 = \gamma_i(p,s) \ket{f_i}
 + \sum_{r=1}^i \hatX_{r,i-r}^{p,s}\,\ket{f_{i-r}}
 \;.
 \label{recur1}
 \ee
By the rules of CFT, since $N_{1,s}\ket{h_{1,s}}$ is a singular vector, the operator product of the field ${\hat{\cN}}_{1,s}\phi_{1,s}(z)$ with any field, or its action on any state, should be ``zero'', so that the states $\ket{\tilde f_i}$ should also be either zero or themselves a null state or a descendant.

The factors $\gamma_i(p,s)$ were found in \cite{BAUER1991515} and are explicitly
\be
\gamma_r(p,s) = \frac{r(p(s-1)-r)}{(s-1)!^2} \prod_{k=1}^{s-2}(kpt - rt - k(s-k-1))
\label{eq:gamma}
\ee
It is clear that $\gamma_0(p,s) = \gamma_{p(s-1)}(p,s)=0$ and so this is consistent with
$\ket{f_i}=0, i<p(s{-}1)$ and 
$\ket{\tilde f_{p(s-1)} }= c\,N_{p,s-1} \ket{h_{p,s{-}1}}$, 
the desired singular vector (up to a normalisation constant $c$).

It is straightforward, then, to solve the recursion relation we get from setting $\ket{f_i}=0, i<p(s-1)$ in equation \eqref{recur1} (omitting the dependence on $(p,s)$ for convenience),
\be
\ket{f_i} = - \frac{1}{\gamma_i}
\sum_{r=1}^i \hatX_{r,i-r}\,\ket{f_{i-r}}
 \;,
 \;\;\;\; i<p(s{-}1)
 \label{recur2}
 \ee
 ie 
\be
\ket{f_i} = 
\sum_{\bm \pi \in \Pi_i}
\prod_{i=1}^{|\bm \pi|}
\left(
- \frac{1}{\gamma_{\tilde\pi_i}} \hatX_{\pi_i,\tilde \pi_i}
\right)
\,\ket{f_0} %{\tilde \pi_i}}
 \;,
 \;\;\;\; i<p(s-1)
 \label{recur3}
 \ee
 where $\tilde \pi_j$ is the partial sum of terms in the partition
 $\bm\pi$ and $l=|\bm\pi|$ is the number of terms in a sequence
 $\bm\pi= (\pi_1,\pi_2,\ldots \pi_l)$, as in \eqref{eq:setdefs}.

This leads to the final expression for the singular vector at level $N=p(s-1)$ in the Verma module $V_{p,s{-}1}$:

\begin{align}
c\,\cN_{p,s-1}\ket{h_{p,s-1}}
=& \ket{\tilde f_{p(s-1)}}
\nonumber\\=&
\sum_{\bm\pi\in \Pi_{p(s-1)}}
\prod_{i=1}^{|\bm\pi|-1}
\left(
- \frac{1}{\gamma_{\tilde \pi_i}}\right)
\cdot 
\prod_{i=1}^{|\bm\pi|}
\left(
\hatX_{\pi_i,\tilde \pi_i}
\right)
\,\ket{h_{p,s-1}}
 \;.
\label{recur4}
\end{align}
This formalism applies for all $p,s$ but so far, to our knowledge, it was only explicitly used for $s=2$, giving the Benoit-Saint Aubin (BSA) formula, for which
\be
\gamma_m(p,2) = t/m(p-m)\;,\;\; \hatX_{m,n}^{p,2} = L_{-m}\;,
\label{gamma1s}
\ee
and for $s=3$ in \cite{Millionshchikov} with correspondingly more complicated formulae.
To extend this to $s>3$, we need to find expressions for the operators
$\hatX_{m,n}$, which we do in the next section. 

\subsection{The general expression of $\hatX_{\pi,\tilde \pi}$}
\label{sec:4}

We start from the BSA form of the state $\cN_{1,s}\ket{(1,s)}$, given by \eqref{recur4} with the substitutions  \eqref{gamma1s}:
\be
\cN_{1,s} \ket{(1,s)} = 
%\left(
\sum_{\bm m\in \Pi_s}\,
\beta_s(\bm m)\, 
L_{-\bm m}
% \prod_{r=1}^{|\bm m|} L_{-m_r}
%\right)
\,\ket{h_{1,s}}
\;.
\label{eq:BSA}
\ee
where $\beta_s(\bm m)$ are defined in \eqref{eq:betadef}.
%and $\tilde m_i$ in \eqref{eq:sumsdef}.
%/we have used the notation
%\be
%\beta(\bm m) = \prod_{j=1}^{|\bm m|-1}
%  \frac{t}{s(s-\tilde m_i)}
%  \;,\;\;
%\tilde m_i = \sum_{j=i+1}^{|\bm m|} m_j
%\;.
%\ee
This means that the field $\hat{\cN}_{1,s}\phi_{1,s}(z)$ takes the form
\be
\hat{\cN}_{1,s}\phi_{1,s}(z)
= 
\sum_{\bm m\in \Pi_s}\,
\beta_s(\bm m)\,
\left(\prod_{r=1}^{|\bm m|} \hat L_{-m_r}\right)
\,\phi_{1,s}(z)
%\right)
\;.
\label{eq:beta}
\ee
For any field $\psi(z)$ of weight $h_\psi$, the operator $\hat L_{-m}\psi(z)$ is defined to  be
\begin{align}
\hat L_{-m}\psi(z)
=& 
\sum_{k\geq 2}
%{\scriptstyle \binom{k-2}{m-2}}
\left({\small\begin{array}{c} k{-}2 \\ m{-}2 \end{array}}\right) 
z^{k-m}  L_{-k}\,\psi(z) 
+ \sum_{k\leq 1} \left({\small\begin{array}{c} k{-}2 \\ m{-}2 \end{array}}\right) z^{k-m} \psi(z)\, L_{-k}
\;,\;\;r>1
\label{eq:Lmode1}
\\
\hat L_{-1}\psi(z)
=&
\frac{\partial}{\partial z}\psi(z)
\;.
\label{eq:Lmode2}
\end{align}
When acting on the state $\ket{h_{p,0}}$, this becomes
\begin{align}
{}&\hat L_{-m}\psi(z)\,\ket{h_{p,0}}
\nonumber\\
=& 
\left(
\sum_{k\geq 2} 
%\left({\scriptstyle{\begin{array}%{c} k{-}2 \\ m{-}2 \end{array}}}
%\right) 
 %
\left({\small\begin{array}{c} k{-}2 \\ m{-}2 \end{array}}\right)  z^{k-m}  L_{-k}\,\psi(z) 
+(-1)^{m}
 \left[ \psi(z) \frac{L_{-1} }{z^{m-1}}
+ \frac{(m{-}1)h_{p,0}}{z^m}  \psi(z) \right]
\right)\,\ket{h_{p,0}}
\nonumber\\
=&
\left(
\sum_{k\geq 1} 
\left({\small\begin{array}{c} k{-}2 \\ m{-}2 \end{array}}\right)  
z^{k-m}  L_{-k}
+ (-1)^m\left(
%\frac {L_{-1}}z   + 
\frac{m h_{p,0} -h_\psi - L_0}{z^m}
 \right)
 \right)\psi(z)
\,\ket{h_{p,0}}
\;, m \geq 2
\label{eq:sub1}
\\
&\hat L_{-1}\psi(z)\,\ket{h_{p,0}}
= \frac{L_0 + h_\psi - h_{p,0}}z\,\psi(z) \ket{h_{p,0}}
\;.
\label{eq:sub2}
\end{align}
where we used the identities true for any scaling field $\psi(z)$ of weight $h_\psi$,
\be
[L_{-1},\psi(z)] = \partial\psi(z)
\;,\;\;
[L_0,\psi(z)] = h_\psi \psi(z) + z \partial \psi(z)
\;.
\ee
We can now apply the rules \eqref{eq:sub1} and \eqref{eq:sub2} successively to each field $\hat L_{-\bm m}\phi_{1,s}$ appearing in \eqref{eq:beta}. If we 
consider a partition $\bm m$ of length $r$, we start with the field
\be
  \hat L_{-m_1}\ldots\hat L_{-m_r}\phi_{1,s}(z)
\;.
\ee
After applying \eqref{eq:sub1} and \eqref{eq:sub2} $(j-1)$ times, we are left with
\be
%  \hat L_{-m_j} \psi_{(m_{j+1},\ldots,m_r)}(z)
%\equiv 
\hat L_{-m_j}\ldots\hat L_{-m_r}\phi_{1,s}(z)
\;.
\label{eq:psidef}
\ee
We can write this as
\be
\hat L_{-m_j} \psi_{(m_{j+1}\ldots m_r)}
\;,
\;\;\;\;
\psi_{(m_{j+1},\ldots,m_r)}(z)
\equiv \hat L_{-m_{j+1}}\ldots\hat L_{-m_r}\phi_{1,s}(z)
\;.
\ee
Recalling that $\tilde m_j = \sum_{q=j+1}^r m_q$, with the choice of $\psi$ as in \eqref{eq:psidef},
%\be\psi(z) = \psi_{(m_{j+1},\ldots,m_r)}\;,
%\ee
we have $h_{\psi} = h_{1,2} + \tilde m_j$ and so, using \eqref{eq:sub1} and \eqref{eq:sub2} again, 
%
%
%If we app
%Each term $\hat L_{-m}\psi(z)$
%will come from some partition $\bm m$ in the sum \eqref{eq:BSA}. If the length of $\bm m$ is r, and we consider the action of $j$-th term $\hat L_{-j}$ then this field will be of the form
%\be
%\psi_{(m_{j+1},..,m_r)}(z) = \hat L_{-m_{j+1}}\ldots \hat %L_{-m_r}\phi_{1,s}(z)
%\;,\;\;
%\sum_{q=j+1}^r m_q = \tilde m_j
%\;,
%\ee
%for which we can write
\begin{align}
&
    \hat L_{-m_j} \psi_{(m_{j+1},..,m_p)}(z) \ket{h_{p,0}}
=
\left(\sum_{k\geq 0}
z^{k-m} S_{m,k} \right)
 \psi_{(m_{j+1},..,m_p)}(z) \ket{h_{p,0}}\;,\;\;
\\
&S_{m,0} = (-1)^m(mh_{p,0} - h_{1,2} - \tilde m_j - L_0)
 \;,\;\;
\label{Sm0}
\\
&S_{m,k} = \begin{cases}
    \left({\small\begin{array}{c} k{-}2 \\ m{-}2 \end{array}}\right)  L_{-k} & m>1,k\geq 1 \;,\\
    0 & m=1,k\geq 1 \;.
\end{cases} 
\end{align}
We can now put all this together to find the contribution of $\psi_{(m_1,\ldots,m_p)}= \psi_{\bm m}$ to the operator $\hatX_{\pi_i,\tilde\pi_i}$. 
The field $\psi_{\bm m}$ contains $r=|\bm m|$ actions $\hat L_{-m_j}$, each of which will contribute a term which lowers the weight of the state by $k_i$ for some $k_i\geq 0$.
Since $X_{\pi_i,\tilde \pi_i}$ lowers the weight by $\pi$, 
we need to sum over all possible choices of 
$k_i$
such that total contribution is to lower the weight by $\pi$, i.e.\ we need to sum over all choices of $k_i$, $1\leq i\leq r$, such that $k_i\geq 0$, $\sum k_i = \pi$: that is we sum over the partitions
\be
\bm k = (k_1,\ldots k_r) \in \Pi_{\pi,r} 
\ee
We can now also calculate the value of $L_0$ in each term $S_{m_i,0}$. This term acts on the state
\be
 S_{m_{i+1},k_{i+1}}\ldots S_{m_{p},k_{p}} \,\ket{f_{\tilde \pi}}
\ee
which has $L_0$ eigenvalue 
\be
h_{p,s-1} + \tilde\pi + \sum_{j=i+1}^r k_j
= h_{p,s-1} + \tilde\pi + \tilde k_j
\ee
Putting this all together, we find an explicit expression for the operator $\hatX_{\pi,\tilde\pi}$ in the formula \eqref{recur4}:
\begin{align}
\hatX_{\pi,\tilde\pi}^{p,s}
 = \sum_{\bm m \in \Pi_s}
 \beta_s(\bm m)
 \sum_{\bm k \in \Pi_{\pi,|\bm m|}}
 \prod_{i=1}^{|\bm m|}
 S^{\tilde \pi,\tilde m_i,\tilde k_i}_{m_i,k_i}
 \;,
 \label{xdef2}
\end{align}
where we have no longer suppressed the dependence of the terms $S_{m,k}$ on $\tilde\pi,\tilde m_i$ and $\tilde k_i$
\begin{numcases}{S^{\tilde\pi,\tilde m,\tilde k}_{m,k} = }
(-1)^m(mh_{p,0} + h_{1,s} -h_{p,s-1}  + \tilde m - \tilde\pi - \tilde k)
 \;,\;\;
 & 
 $k=0\;,$ 
\label{eq:Smo3}
\\[2mm]
 0 \;,\;\; & 
 $k>0,m=1 \;,$
 \\[2mm]
 \binom{k-2}{m-2}
 L_{-k}\;, 
 & $k>0,m>1\;.$
\label{eq:Smk3}
\end{numcases} 
%\end{align}
  %
\blank{
\begin{align}
&S^{\tilde\pi,\tilde m,\tilde k}_{m,0} = (-1)^m(mh_{p,0} -h_{p,s-1} - h_{1,s} + \tilde m - \tilde\pi - \tilde k)
 \;,\;\;
 \label{eq:Smo}
 \\
&S^{\tilde\pi,\tilde m,\tilde k}_{m,k} = \begin{cases}
    \left({\small\begin{array}{c} k{-}2 \\ m{-}2 \end{array}}\right)  L_{-k} & m>1,k\geq 1 \;,\\
    0 & m=1,k\geq 1 \;.
\end{cases} 
\label{eq:Smk4}
\end{align}
}
This, together with the definitions of the factors $\beta_s(\bm m)$ \eqref{eq:betadef} and $\gamma_r(p,s)$ \eqref{eq:gamma}, completes the explicit from of the singular vector in \eqref{recur4}, up to the constant $c$. Since $\cN_{p,s-1} = (L_{-1})^{p(s-1)} + \ldots$, we see that $c$ is simply the coefficient of $(L_{-1})^{p(s-1)}$ in 
\eqref{recur4}, which we denote by $\zeta_{p,s}$ and for which we find an expression in the next section.

As a final comment, it is easy to recover $X_{m,n}^{p,2}=L_{-m}$
in \eqref{gamma1s} from \eqref{xdef2}.

\section{Derivation of the second formulation}
%coefficients $\zeta({\bm m})_{p,s}$.}
%The general expression of $\cN_{r,s-1}$ (second version)}
\label{sec:5}

We have already seen \eqref{eq:BSA}
that the formula of BSA can also be written as a sum of an (over-complete) set of products of modes. We now show how to rearrange the result \eqref{recur4} in this form,
\be
c\,\cN_{p,s-1} 
= \sum_{{\bm m}\in\Pi_{p(s-1)}} \zeta(\bm m)_{p,s}\, \prod_{m_i\in\bm m} L_{-m_i}
\;,
\label{eq:zeta1}
\ee
by finding explicit formulae for the coefficients $\zeta(\bm m)$.
The method is simple: each factor $L_{-m_i}$ must come from some operator $\hatX_{\pi,\tilde\pi}$, , and each operator $\hatX_{\pi,\tilde\pi}$ can contribute from 1 to $s$ factors $L_{-m}$.
Accordingly, we split the product of modes $L_{m_i}$ into sub-products, each of which comes from one operator $\hatX$. 
We then calculate the coefficient for each of these and sum them up. The normalisation constant $c$ is then just $\zeta_{p,s}=\zeta((1,1,..1))_{p,s}$.% and so we can find the precise form of $\cN_{p,s-1}$, \eqref{eq:Nps}.

We now consider a single subsequence $\bm m$ in \eqref{eq:zeta1}. We split this into $l$ subsequences of lengths $k_j$, so that $\bm k$ is a partition of $|\bm m|$ with $|\bm k|=l$, as in figure \ref{fig:1}.
%\begin{landscape}
\begin{figure}[htb]
\[   
\begin{array}{c}
\overbrace{
\vphantom{\Biggr(}
\underbrace{
L_{-m_1} \;\;\ldots\;\; L_{-m_{k_1}}
}_{\text{$k_1$ terms}}
\;\cdots\;
\underbrace{
L_{-{m_{k_1 +\ldots+k_{j-1}+1}}}\;\;\ldots\;\; L_{-m_{k_1+\ldots k_j}}
}_{\text{$k_j$ terms from
$X_{(\tilde m_{\hathat k_{j-1}}{-}\tilde m_{\hathat k_j}),\tilde m_{\hathat k_j}}$}}
\;\cdots\;
\underbrace{
L_{-m_{n-k_l+1}} \;\;\ldots\;\; L_{-m_n}
}_{\text{$k_l$ terms, $l=|\bm k|$}}
}^{\text{$n = |\bm m|$ terms corresponding to the partition $\bm m$ in $\zeta(\bm m)$}}
\\
\overbrace{
L_{-{m_{\hathat k_j + 1}}}
\qquad
\;\;\ldots\;\; 
\qquad
L_{-m_{\hat k_j }}
}\phantom{aaa}
\end{array}
\]
\caption{Splitting the partition $\bm m$ into $l$ subsequences. 
Each of the terms from $j=2$ to $j=l$ carries an overall factor of \hbox{$(-1/\gamma_{\tilde m_{\hat{\hat k}_j}}(p,s))$}
}
\label{fig:1}
\end{figure}
Each of the $l$ subsequences will come from a single operator $X$, and we can work out that the $j$-th subsequence comes from the operator $X_{\tilde m_{\hathat k_{j-1}} - \tilde m_{\hathat k_j},\tilde m_{\hathat k_j}}$. This means the contribution from the division of $\bm m$ into these $l$ subsequences carries an overall factor of
\be \prod_{j=2}^{|\bm k|}
(-1/\gamma_{\tilde m_{\hathat k_j}}(p,s))
\;,
\ee
which is the first factor in \eqref{eq:zeta2}.
%$
%se are . The 
%, and so comes with an overall 

We now restrict attention to the $j$-th subsequence, corresponding to the product of models
\be 
\prod_{i=\hathat k_j + 1}^{\hat k_j}
L_{-m_l}
= L_{-m_{\hathat k_j+1}} \cdots L_{-m_{\hat k_j}}
\;.
\label{eq:kprod}
\ee
There are $k_j$ modes $L_{-m_{\ldots}}$ in this product, and these can come from any term in \eqref{xdef} which has enough $\hat L$ terms. Of these, $k_j$ of the $\hat L$ terms will give a lowering mode $L_{-m_{\ldots}}$, and the rest will just give a constant factor. 

Suppose that the $k_j$ modes \eqref{eq:kprod} come from a term in
\eqref{xdef} corresponding to a partition $\bf q$ (with overall factor
$\beta_s(\bm q)$). We need to assign $k_j$ of the terms in $\bf q$ to
a lowering operator $L_{-m_{...}}$ using \eqref{eq:Smk} and the
remaining $(|\bf q|-k_j)$ modes simply give a constant factor,
according to \eqref{eq:Smo}.  One way to describe this is with (yet
another) partition, splitting the $(|\bf q| - k_j)$ modes into exactly
$k_j{+}1$ sets of length $v_w$ where $\bf v\in\Pi_{|\bf
  q|-k_j,k_j+1}$, as shown in figure \ref{fig:2}. Inside each group of
$v_w$ term, we will use an index $x$ to label the terms in our final
expression.
 %where the $w$-th set appears between the $w{-}1$-th and $w$-th modes $L_{-m...}$.
 (This could equally well be described by an increasing sequence $1 \leq i_1 < i_2..< i_{k_j} \leq |\bm q|$ where $i_w = \hat v_w$, but we will stick to the partition notation).
%This is shown in figure \ref{fig:2}.

%\begin{landscape}
\begin{figure}[htb]
\[   
\scalemath{0.95}{
\begin{array}{c}
\underbrace{
\begin{array}{rccccccl}
&\multicolumn{6}{c}{{\overbrace{\phantom{aaaaaaaaaaaaaaaaa AAAAAAAA AAAAAAA  AAAAA AAA AAAAA aaaa}}^{\text{$k_j$ modes $L_{-m_{\ldots}}$}}}}
\\
{\text{const.}\times (}
& L_{-m_{\hathat k_j+1}}
&
\cdots L_{-m_{\hathat k+j+w-1}}
&\cdots&
L_{-m_{\hathat k_j+w}}&\cdots&L_{-m_{\hat k_j}}&)\\
& \uparrow &\uparrow&&\uparrow&&\uparrow\\
\underbrace{\phantom{a}\cdots\phantom{a}}_{\text{$v_1$ factors}}\, 
& 
\, L_{-m_{\hathat k_j + 1}}
& \cdots
\, L_{-m_{\hathat k_j + w-1}}
& 
\underbrace{\phantom{aaaaaaaaa}\cdots\phantom{aaaaaaaaa}}_{\text{$v_w$ constant factors}}\, 
& 
\, L_{-m_{\hathat k_j + w}}
& \cdots
%& 
%\underbrace{\phantom{aa}\cdots\phantom{aa}}_{\text{$v_{k_j}$ terms}}\, 
&
\; L_{-m_{\hat k_j}}
\;
& 
\underbrace{\phantom{a}\cdots\phantom{a}}_{\text{$v_{k_j{+}1}$ factors}}\, 
\\
&&\uparrow & \uparrow\hbox to 0pt{${}^{**}$} & \uparrow\hbox to 0pt{${}^*$} \\
&\cdots&
\hat L_{-q_{\hat v_{w-1} + w-1}}
&
%\underbrace{
\overbrace{
{
\underset{\scriptscriptstyle x=1}{\hat L_{-q_{\hathat v_w + w}}}}\cdots
\underset{\scriptscriptstyle x=v_w}{\hat L_{-q_{\hathat v_w + v_w + w-1}}
}}^{\text{$v_w$ terms}}
%}_{\text{$v_w$ terms}}
&
\hat L_{-q_{\hat v_w + w}}
&&\cdots
\\[2mm]
%&&&x=1\cdots x=v_w & \\
%\multicolumn{7}{c}{{\underbrace{\phantom{AAAA}}_{\text{terms arising from $\hat L_{-q_1}\ldots \hat L_{-q_{|\bm q|}} \phi_{1,s}$}
%}}}
\end{array}
}_{\text{terms arising from $\hat L_{-q_1}\ldots \hat L_{-q_{|\bm q|}} \phi_{1,s}$}
}
\end{array}}
\]
\caption{{Assigning $k_j$ of the $|\bm q|$ terms $\hat L_{-q_i}$ to operators $L_{-m_n}$ and the rest to constant factors. The term marked ${}^*$ gives the factor \eqref{eq:*factor}; the terms marked ${}^{**}$ give the factors \eqref{eq:**factor}
for $1\leq x\leq v_w$.
}}
\label{fig:2}
\end{figure}
To work out the constant in figure \ref{fig:2}, we will need to work out the contribution of each constant term, and the factor for each mode $L_{-k}$.

The factor for each mode $L_{-k}$ is easy - it is either a binomial if it comes from $\hat L_{-m}$ with $m>1$, or zero, if it comes from $\hat L_{-1}$, and we denote this factor by
\begin{numcases}{S_0(m,k)= }
 0 \;,\;\; & 
 $m=1 \;,$
 \\[2mm]
 \binom{k-2}{m-2}
 \;, 
 & $m>1\;.$
\label{eq:S3b}
\end{numcases} 
For the $w$-th term in the $j$-th subsequence, marked $*$ on figure 2, altogether this is the $(k_1 +\ldots k_{j-1} + w)$-th mode, i.e.\ the mode is $L_{-m_{\hathat k_j + w}}$, and it arises from $\hat L_{-q_{\hat v_w + w}}$, and so the factor is 
\be
S_0(q_{\hat v_w + w},m_{\hathat k_j + w})
\;.
\label{eq:*factor}
\ee

For a constant term arising from $\hat L_{-q}$, we need to know both the level of the state on which it acts, and the level of the descendent field on which $\hat L_{-q}$ acts.
If the operator $\hat L_{-q}$ is in the $w$-th group, then it acts on a state at level ${\tilde m}_{\hathat k+j + w - 1}$; if it is the $n$-th term in the partition $\bm q$, it comes from the action on a descendent field at level $\tilde q_n$.
Furthermore, if it is in the $w$-th group, $n$ must be equal to $\hathat v_w + w + x - 1$ for some $x$ in the range $1 \leq x \leq v_w$.
If
we denote the constant factor arising from the action $\hat L_{-r}$ on a descendent field of level $k$, acting on a state at level $m$, by
\begin{align}{S_1(r,k,m)= }
(-1)^r(k+rh_{p,0}+h_{1,s}-h_{p,s-1} - m)
 \;,\;\;
 & 
 \label{eq:S3a}
\end{align} 
then the factor from the $x$-th term $\hat L_{-q_{\ldots}}$ in the $w$-th group, the factor is 
\be
 S_1(q_{\hathat v_w + w + x - 1},\tilde q_{\hathat v_w + w + x - 1},
    \tilde m_{\hathat k _j + w - 1})
\;.
\label{eq:**factor}
\ee
As can be seen, the terms from the $j$-th subsequence only depend on that subset of modes, and so it is helpful to denote these modes by the sequence $\bm m^j$,
%the $j$-th subsequence of $\bm m$,
\be
\bm m^j = (m_{\hathat k_j+1},\ldots,m_{\hat k_j})
\;,
\ee
and then
\be
{\tilde m}_{\hathat k_j + w}
= {\tilde m}^j_{w} + \tilde m_{\hat k_j}
\;.
\ee
Summing over all the partitions $\bm q$, and noting that $k_j = |\bm m^j|$, we obtain the expression for the contribution from all the states in $\cN_{1,s}\ket{h_{1,s}}$ to the 
terms $L_{-\bm m^j}$ which corresponds to the 
$j$-th subsequence of $\bm m$,
\begin{align}
&Y(\bm m^j,\tilde m_{\hat k_j})
= 
\sum_{\bm q \in \Pi_s}
\beta_s(\bm q)
\Biggl[
\nonumber\\
&
\sum_{\bm v \in \Pi_{|\bm q| - k_j,k_j}}
\left(\prod_{w=1}^{k_j}
S_0(q_{\hat v_w + w},m^j_w))\right)
\left(
\prod_{w=1}^{k_j+1}
\prod_{x=1}^{v_w}
S_1( q_{\hathat v_w +w + x -1} ,
{\tilde q}_{\hathat v_w + w _ x -1},
{\tilde m}^j_{w-1} + {\tilde m}_{\hat k_j})
\right)\Biggr]
\;,
\end{align}
where
$ k_j = |\bm m^j| $.
Putting the various sums over partitions and these factors together, we end up with the expressions given already in section \ref{sec:f2}.

\section{Conclusions and Outlook}
\label{sec:conc}

We have found two (equivalent) expressions for the singular vector
$\cN_{r,s}$, \eqref{eq:f1} and \eqref{eq:f2}. These are both
generalisations of the expressions found by Benoit and Saint Aubin for
$\cN_{1,s}$. These are of course very far from unique. Quite apart
from the fact that the set of vectors in \eqref{eq:zeta1} are
over-complete, the original paper of Bauer et al \cite{Bauer:1991qm}
indicated how the ```fusion point'' would lead to different
expressions, as was shown in \cite{Bowcock:1992gt} where expressions
for $\cN _{1,s}$ were found using only the modes $L_{-1}$ and $L_{-2}$.
The proof that these expressions really do define singular vectors is contained in \cite{BAUER1991515}.
As noted in that paper, the algorithm does not respect the symmetry
$r\leftrightarrow s, t \leftrightarrow1/t$, but the results do. 

We should note that the factors $\gamma_n$ have zeroes for certain values of $t$ for $s>2$, so that formally these expressions are not defined for all $t$ since they contain terms $-1/\gamma_n$.
The zeroes in $\gamma_n$ correspond to situations where the singular vector operator $\cN_{p,s-1}$ factorises, i.e. the singular vector $\cN_{p,s-1}\ket{h_{p,s-1}}$ can be written as
\be
  \cN_{p,s-1}\ket{h_{p,s-1}} = \cN_{r_1,s_1} \cN_{r_2,s_2} \ldots \cN_{r_p,s_p} \ket{h_{p,s-1}}
  \;,
\ee
Since there is only ever one singular vector at any particular level $N$ \cite{Fuchs}, and this is uniquely characterised by the coefficient of $L_{-1}^N$, the poles in $\gamma_n$ are cancelled corresponding zeroes in $\zeta((1,1,\ldots,1))$, and the resulting expressions will be regular for $t$ non-zero.

It would of course be nice to find a simpler formula than \eqref{eq:zeta2} and this might seem easy as the results in \cite{BENOIT1988517} and \cite{Millionshchikov} are indeed much simpler, each term $\zeta(\bm m)$ being a product of factors which are linear in $t$. On ``experimental'' evidence, $\zeta((1,1,...,1))$ is also a product of linear factors in $t$ times a non-positive power of $t$, 
\be
\zeta_{p,s} = \zeta((1,...1))_{p,s}
= (-1)^{ps-1} ((p-1)!(s-1)!)^2 t^{2p - ps - 1}\prod_{i=1}^{p-1}\prod_{j=1}^{s-1}
(it-j)(it+j)
\ee
and it should certainly be possible to prove, this form - but not all coefficients seem to factorise over $\Bbb Q$, for example the coefficient $\zeta((7,2))_{3,4}$ in $\cN_{3,3}$ is 
\be
\zeta((7,2))_{3,4}
= \frac{5 \left(5 t^2-16 t-9\right)}{2 (t-2) (2 t-1)}
\;,
\ee
and it seems polynomials of arbitrarily high can appear in $\zeta$,
for example $\zeta((3,3,1,2))_{3,4}$ has a factor of order 6. The
functions $Y$ defined in \eqref{eq:Ydef1} also do not factorise in general.

We should also note that the number of states in the expansion
\eqref{eq:zeta2} grows much faster with level than the dimension of
the Verma module. For example,  the subspace of the
Verma module of level 16 has
dimension 231, which is the number of ordered partitions of 16, but
the number of unordered partitions is $2^{15} = 32,768$, so that there
are 231 terms in the expression of $\cN_{4,4}$ in the form
\eqref{eq:psinorm} but 32,768 terms in the expression of the same
operator $\cN_{4,4}$ in the form \eqref{eq:f2}. As a consequence, the
new expressions presented here seem to have little practical utility in general.  It is hoped,
however, that at least having the formula \eqref{eq:zeta2} will help
find a simple form for $\zeta((\bm m))$.

Finally, there is a Mathematica notebook \cite{SV} available which
implements the algorithms presented here and includes an
implementation of the Virasoro algebra by Matthew Headrick
\cite{Headrick} to enable some checking of the results.

\blank{
For example, 
. However, the zeroes in $\gamma_n$ correspond to situations where the singular vector operator $\cN_{p,s-1}$ factorises, i.e. the singular vector $\cN_{p,s-1}\ket{(p,s-1)}$ can be written as
\be
  \cN_{p,s-1}\ket{(p,s-1)} = \cN_{r_1,s_1} \cN_{r_2,s_2} \ldots \cN_{r_p,s_p} \ket{(p,s-1}
  \;,
\ee
The simplest case when this occurs is at $t=1$ when $h_{1,1} = h_{2,2} = 0, h_{1,3} = 1$, so that using the formulae in this note we find 
\be
\cN_{2,2} = 

\be
\lim_{t\to 1} \left(  \cN_{1,3} \,\cN_{1,1} \right) = \lim_{t\to 1} \left(
\frac{-t(t^2-1)}4 \cN_{2,2} 
\right)
\ee
}

\newpage
\section*{Acknowledgments}
The expressions in section \ref{sec:3} were actually found in 1992,
when the author was supported by St John's College, Cambridge.  GW is
very grateful to Adrian Kent, Matthias Doerrzapf and Peter Goddard for
discussions at this time. The expressions in sections \ref{sec:5} are
recent. GW would like to thank S. Ribault and I. Runkel for
discussions, and the Mathematics Department, University of Hamburg,
for hospitality while this work was completed.  GW is grateful for
support from the DFG (German Research Foundation) via the Cluster of
Excellence EXC 2121 “Quantum Universe” - 390833306, and via the
Collaborative Research Centre CRC 1624 “Higher structures, moduli
spaces and integrability” - 506632645, and from the STFC for support
under grant ST/T000759/1.

\bibliographystyle{JHEP}
\bibliography{Notes}

\end{document}